\documentclass[12pt]{article}
\usepackage{jheppub}

\def\CN {{\cal N}}

\title{Kazama-Suzuki models and BPS domain wall junctions in $\CN=1$ $SU(N)$ Super Yang-Mills}

\author[1]{Davide Gaiotto}

\affiliation[1]{Perimeter Institute for Theoretical Physics, 
31 Caroline Street North, ON N2L 2Y5, Canada}

\abstract{Domain walls in $\CN=1$ supersymmetric Yang-Mills theory conjecturally support topological degrees of freedom 
at low energy. Domain wall junctions are thus expected to support gapless degrees of freedom. We propose a natural candidate for the low-energy description of such junctions. 
}

\begin{document}

\maketitle

\section{Introduction}

Four-dimensional $\CN=1$ $SU(N)$ pure gauge theory has $N$ massive vacua, which break spontaneously 
the $Z_{2N}$ R-symmetry of the theory down to a $Z_2$ subgroup which acts trivially on all bosonic operators. 
See \cite{Weinberg:2000cr} for a useful review and further references.
The vacua are distinguished by the vev of the gaugino condensate, which is 
\begin{equation}
\langle \lambda \lambda \rangle = e^{\frac{2 \pi i k}{N}} \Lambda^3
\end{equation}
in the $k$-th vacuum of the theory. 
The superpotential $W_k$ associated to the $k$-th vacuum is also proportional to $e^{\frac{2 \pi i k}{N}} \Lambda^3$.
\begin{figure}
\center
\includegraphics[width=3.5in]{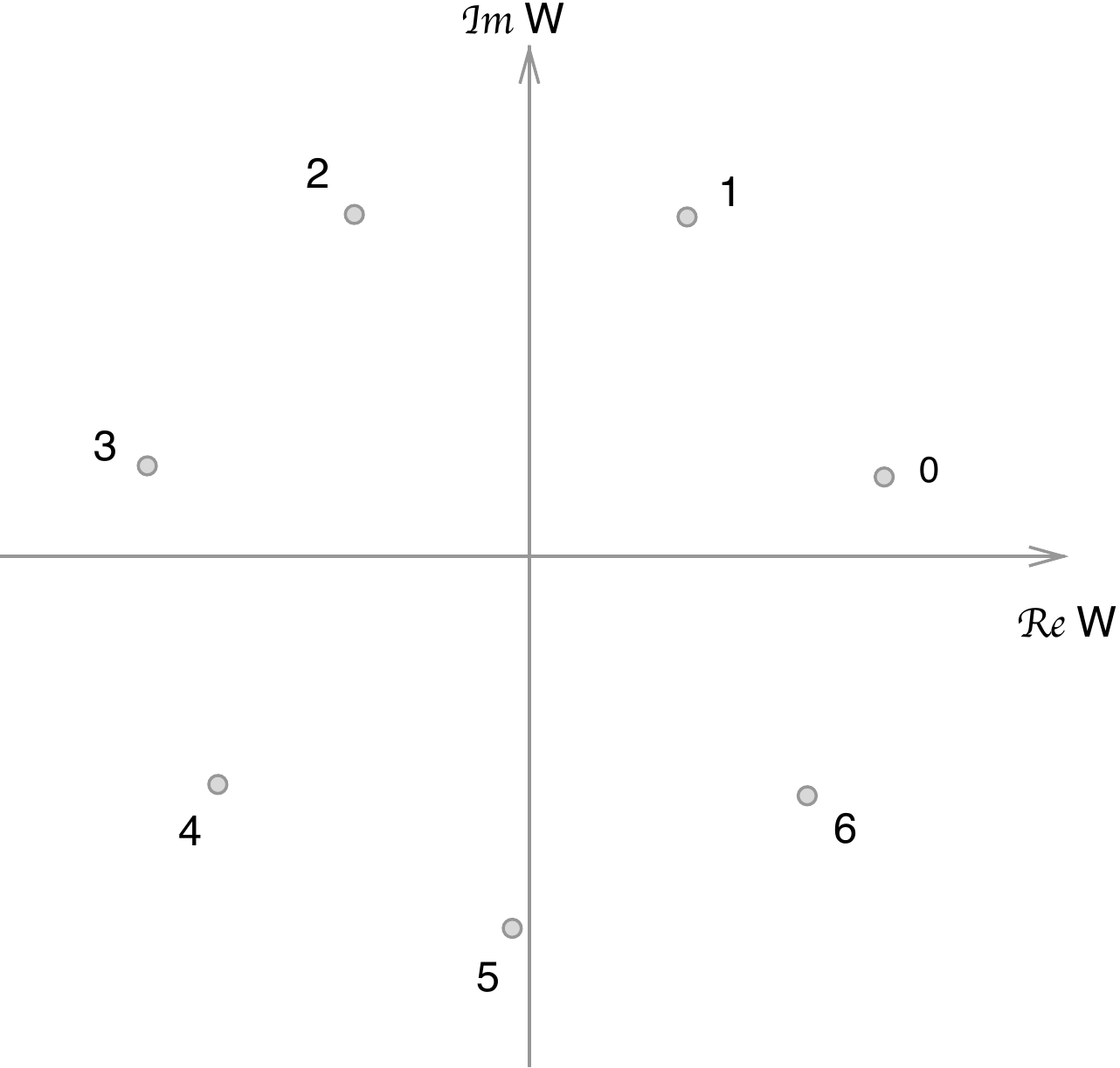}
\caption{The values of the super potential in the seven vacua of $\CN=1$ $SU(7)$ gauge theory}
\label{fig:one}
\end{figure}

Conjecturally, the theory admits a dynamical half-BPS domain wall between each pair of vacua. 
The domain wall between vacua $i$ and $j$ has central charge $W_{ij} = W_i - W_j$ and tension $|W_{ij}|$. 
Due to the $Z_{2N}$ R-symmetry, the properties of the wall depend only on the phase difference $i-j$ modulo $N$. 
As the domain wall breaks translation invariance, it will support a translational zero-mode, consisting of a single 
three-dimensional $\CN=1$ real chiral super-field. 

It is natural to wonder what other low-energy degrees of freedom, if any,
may be supported on the wall. Although it is likely that no other gapless modes are supported on the wall,
the proposal of \cite{Acharya:2001dz} strongly suggests that the domain walls will support a non-trivial 3d TFT
at low energy, a supersymmetric Chern-Simons theory. 
\begin{figure}
\center
\includegraphics[width=3.5in]{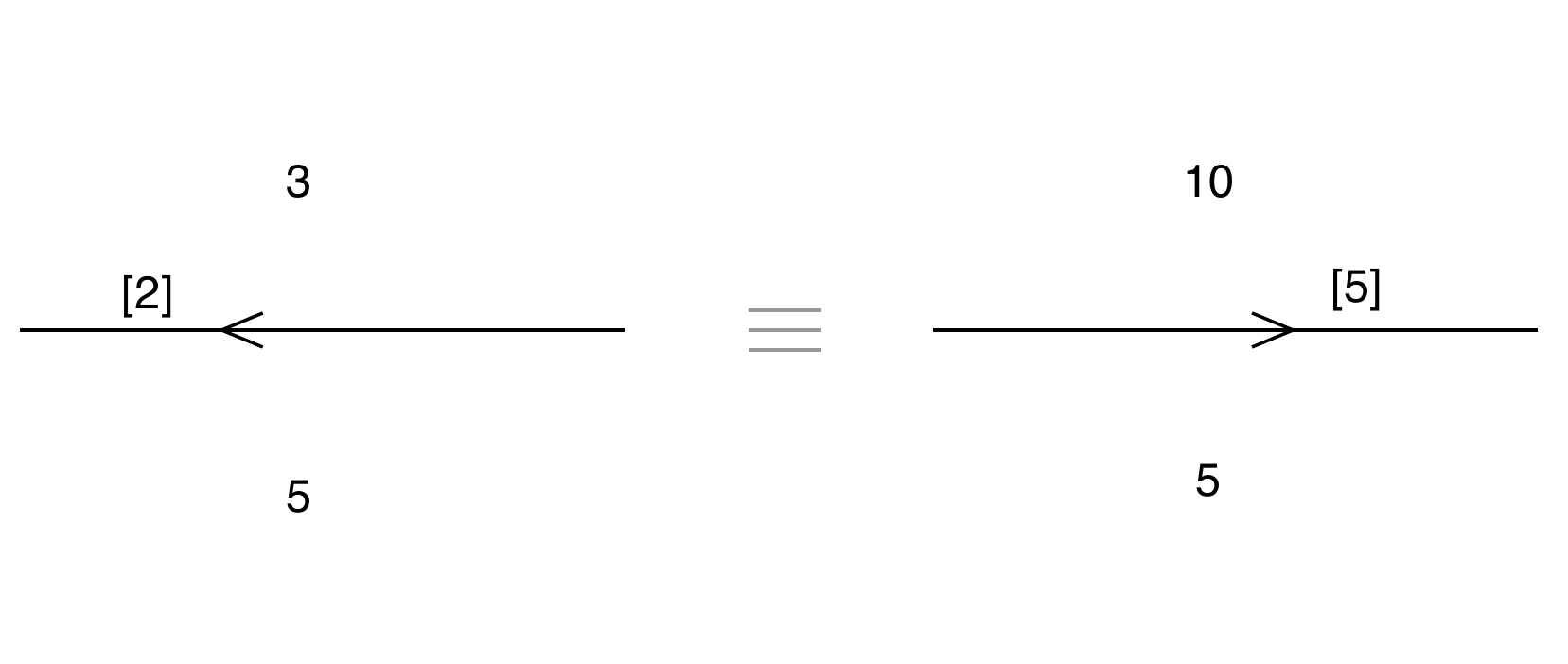}
\caption{A domain wall between the third and fifth vacua of $\CN=1$ $SU(7)$ gauge theory. We label the domain wall by the difference $5-3=2$. The same domain wall, with opposite orientation, 
can be labelled by the difference $3-5$ modulo $7$, i.e. $5$. The label of a domain wall determines the low energy 3d TFT supported on the wall.}
\label{fig:two}
\end{figure}

Whenever one has BPS domain walls in a $\CN=1$ theory, it is natural to look at quarter-BPS junctions between domain walls \cite{Gibbons:1999np,Carroll:1999wr,Townsend:1999nf}. 
A quarter BPS network of domain walls is a planar web configuration, with a vacuum assignment in each 
face (finite or infinite) of the web and wall segments between vacua $i$ and $j$ 
oriented in the plane with a slope related to the phase of their central charge, as in figure \ref{fig:three}. 
The configuration preserves 2d $(1,0)$ supersymmetry. 
Considerations based on effective low energy models \cite{Gorsky:1999hk,Gabadadze:1999pp} suggest that $\CN=1$ $SU(N)$ pure gauge theory should admit such junctions. 

Because of the relation between the slopes of domain walls and of the superpotential differences $W_{ij}$, 
the sequence of vacua around a BPS junction is ordered in the same way as the superpotentials of the vacua around the unit circle. 
Up to a $Z_{2N}$ rotation, we can characterize every possible junction by a cyclically ordered sequence of positive integers $n_a$ which add to $N$, 
so that the sequence of vacua around the junction is $\ell$, $\ell+n_1$, $\ell+n_1 + n_2$, $\ell+n_1 + n_2 + n_3$, etcetera. 
See figure \ref{fig:three} for a simple example. 
\begin{figure}
\center
\includegraphics[width=3.5in]{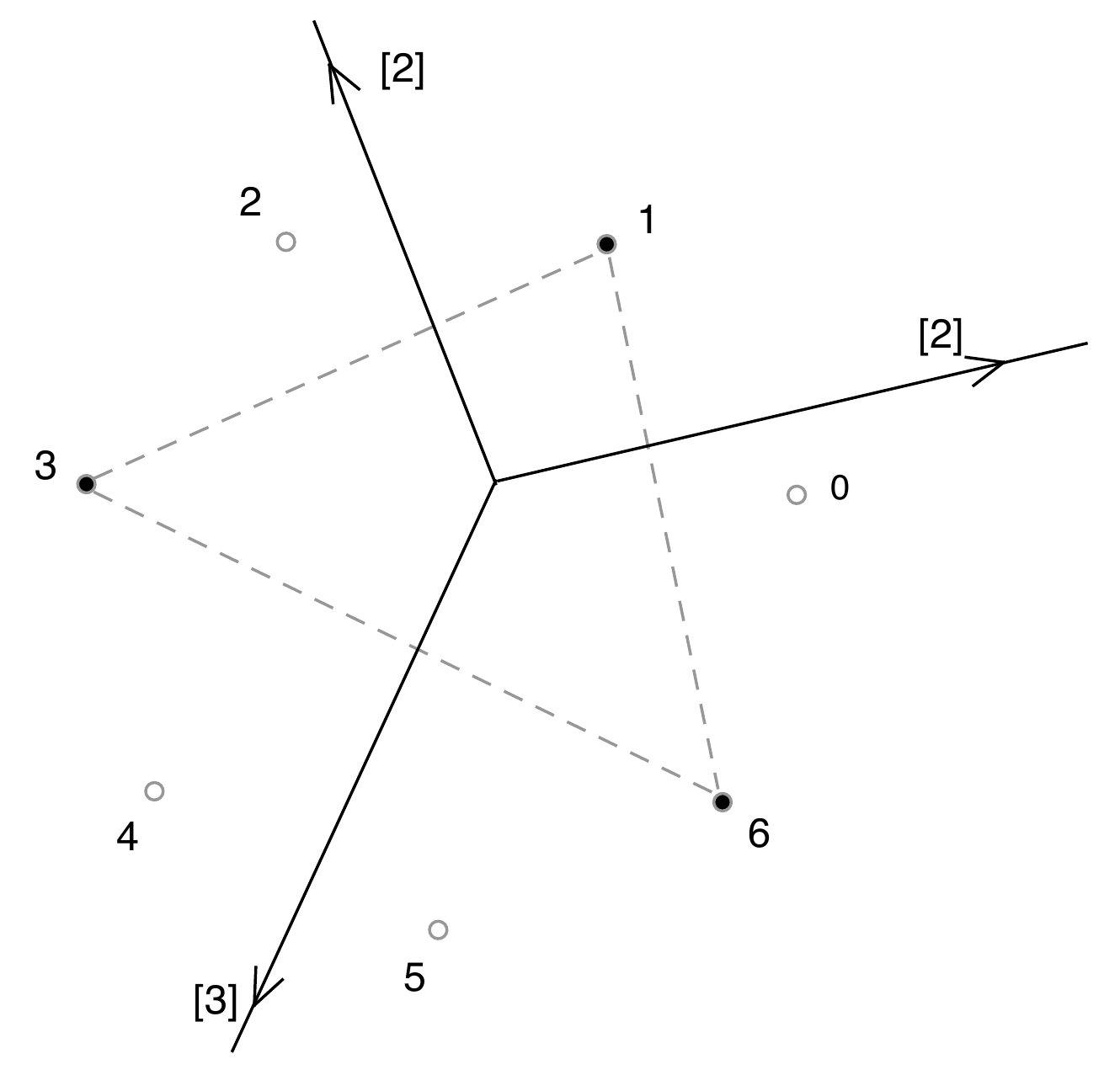}
\caption{A junction of domain walls between the first, third and sixth vacua of $\CN=1$ $SU(7)$ gauge theory.
We superimpose the superpotential plane to the picture of the junction, to highlight the relation between the slope of the walls and
 the phase of the superpotential differences $W_{ij}$.
This relation constrains the vacua around the junction to be cyclically ordered according to their position in the superpotential plane.  
This junction can be labelled by the cyclic sequence $2,3,2$, the labels of the domain walls ending at the junction. We denote as black dots the vacua in each region. }
\label{fig:three}
\end{figure}

As the domain walls support non-trivial 3d TFTs, it is likely that the junctions will have to support appropriate gapless chiral 
degrees of freedom \cite{Ritz:2006zz}. The purpose of this note is to propose candidate low energy world volume theories for such junctions:
the supersymmetric WZW chiral coset models 
\begin{equation}
T_{(n_a)}= \left[ \frac{u(N)_N}{\prod_a u(n_a)_N} \right]
\end{equation}
Our proposal is based on two basic consistency requirements: the junction theory should be a $(1,0)$ SCFT which 
couples naturally to the domain walls 3d TFTs and behaves smoothly under manipulations of the junctions which preserve the quarter-BPS 
condition. 

Clearly, the correctness of our proposal hinges crucially on the correct identification of the domain walls 3d TFTs.
If our assumption about the nature of these 3d TFTs are incorrect, the identification of the junction theories will be incorrect as well. 
We will discuss the soundness of our assumption at some length in the body of the paper. On the other hand, the 
 ``junction splitting'' consistency condition we impose on the junction theories appears to be rather strong. 
 We consider the existence of a candidate family of junction theories compatible with the junction splitting condition to be 
 a non-trivial consistency check on our initial assumption.
 
The theories $T_{(n_a)}$ are actually a Kazama-Suzuki coset models and thus possess a hidden $(2,0)$ supersymmetry. 
It is unclear to us if this fact should be considered a coincidence, or should have a deeper meaning. 
Perhaps this low-energy enhancement of supersymmetry can be motivated by the observation that a rotation in the plane of the 
junction changes the subset of bulk supercharges preserved by the junction, but it really only acts on the domain walls, which are topological at low energy. Thus a junction may naturally acquire at low energy the maximum amount of supersymmetry for a 
codimension two defect in a $\CN=1$ theory, which is $(2,0)$. 

In section \ref{sec:one} we will review the conjectural domain wall theories and present the basic consistency checks for our proposal. 
In section \ref{sec:two} we will review super-WZW and super-coset models and present some details of the interaction of the
line defects in the domain wall theories with the operators of junction theories. 
In section \ref{sec:three} we will offer an alternative perspective on the problem, based on the mass deformation of $\CN=2$ theories.
We will conclude with a discussion of possible generalizations of our proposal.
\section{The basic construction} \label{sec:one}

The authors of \cite{Acharya:2001dz} propose a candidate world volume theory for the BPS domain walls of $\CN=1$ $SU(N)$ SYM: 
a domain wall between the $\ell$-th and the $(\ell+n)$-th vacua (modulo $N$)
is expected to support a $\CN=1$ $U(n)$ Yang-Mills Chern-Simons theory, coupled to a massless adjoint real chiral field. 
The diagonal part of the adjoint chiral field coincides with the translation zero-mode. 
The rest of the theory will presumably be gapped at low energy: 
no non-renormalization theorem prevents the traceless adjoint chiral field from getting a real superpotential. 

If we assume that the adjoint chiral field simply receives a mass and does not develop a vev in the IR, 
we can integrate away the massive degrees of freedom and find at low energy a
topological $\CN=1$ $U(n)$ Chern-Simons gauge theory at level $N$. 
This is the 3d TFT we will assume to arise at low energy on the BPS domain walls of 
$\CN=1$ $SU(N)$ SYM.

It is possible to consider an alternative scenario, where the adjoint chiral field receives a vev and 
thus Higgses the 3d gauge theory down to a Chern-Simons theory with a smaller gauge group. 
Such alternative scenario would possibly lead to a disagreement with the expected Witten index ${N \choose n}$ for the 
domain wall theory on a two-torus (with the centre of mass zero-mode removed) and thus require, say, a phase transition as 
the size of the torus is changed. 

As a further cautionary note, it is useful to point out that the string theory construction of \cite{Acharya:2001dz} 
does not strictly imply that at some intermediate energy scale the domain walls of $\CN=1$ $SU(N)$ SYM
can be described by the Chern-Simons theory coupled to the massless adjoint. 
The regime of parameters where the bulk string setup produces a decoupled $\CN=1$ $SU(N)$ SYM
differs from the regime of parameters where the 3d gauge theory description of domain walls is trustworthy. 
Thus we must assume that the low energy 3d TFT is sufficiently robust to remain unchanged as the parameters 
are tuned from one regime to the other. In other words, we must assume the absence of a phase transition.
We will elaborate on this point further in section \ref{sec:three}.

\subsection{Coupling to the junctions}
A $\CN=1$ supersymmetric Chern-Simons theory differs from a bosonic Chern-Simons theory by the addition of 
an adjoint fermionic auxiliary field. If regulated by a Yang-Mills action, the fermionic field is a standard 
real, massive fermion, with a mass term proportional to the Chern-Simons level. The topological effect of the extra massive fermion
can be seen in the natural choice of a BPS boundary condition for the $\CN=1$ Chern-Simons theory: coupling to
a two-dimensional $(1,0)$ supersymmetric (anti)chiral $[u(n)]_N$ WZW model with level $N$ for the total WZW currents.

A supersymmetric WZW model consists of a bosonic WZW model together with a set of real fermions in an adjoint representation of the group. The total currents are the sum of the bosonic currents and a fermion bilinear, and the total level $k=N$
of these currents is the sum of the bosonic level $\hat k$ and the level of the fermionic currents, which is $n$ for $su(n)$. 
For example, the $[u(n)]_N$ supersymmetric WZW model with level $N$ is built from a level $N-n$ bosonic model and  
the parameter $n$ lies necessarily in the range $1 \leq n \leq N$. In particular, the $[u(N)]_N$ theory, which consists of $N^2-1$ free fermions plus the diagonal $[u(1)]$ factor, has a single NS primary field and thus can be coupled to a somewhat trivial 3d TFT. 

The chiral super-coset 
\begin{equation}
T_{(n_a)}= \left[ \frac{u(N)_N}{\prod_a u(n_a)_N} \right]
\end{equation}
is an RCFT with primaries labelled by a collection $(\lambda_a)$ irreducible representations of the 
$[ u(n_a)]_N$ current algebras. The braiding and fusion matrices will coincide with the ones for an anti-chiral 
$\prod_a \left[u(n_a) \right]_N$ super-WZW model. In particular, this super-coset can be coupled to the 
$U(n_a)_N$ supersymmetric CS theories living on the world volumes of the domain walls, as desired. 
\footnote{It is important to observe that 
this proposal is consistent only for a specific choice of sign of the Chern-Simons couplings on the domain wall theories.
The choice of supersymmetry preserved by the junction determines both the chirality of the $(1,0)$ supersymmetry algebra 
and the orientation of the walls around the junction, which together with the sign of the CS coupling determines 
if the junction theory should behave as a chiral or an anti-chiral $\left[ \prod_a u(n_a)\right]_N$ super-WZW model.
Our proposal would be inconsistent if the sign of the CS coupling required a $(1,0)$ junction theory behaving as a chiral 
super-WZW model rather than the chiral coset $T_{(n_a)}$. We could not find solutions to the junction splitting constraints 
which are consistent with the alternative choice of 
signs for the Chern-Simons couplings. The careful analysis required to fix the precise relative sign predicted by the 
brane model goes beyond the scope of this paper and we leave it for future work. }

\subsection{Junction manipulation}
Any junction between four or more domain walls can be split in several distinct ways into a web with two junctions, joined by a finite wall segment. The 2d theory living on the original junction and the product of the two theories living at the two new junctions might possibly be different, but they must have the same gravitational anomaly.
Thus if we spit the sequence $(n_a)$ into $(n'_a)$ and $(n''_a)$, we must find 
that the theory $T_{(n_a)}$ and the product theory 
\begin{equation}
T_{(n'_1, n'_2, \cdots, \sum_a n''_a)} \times T_{(n''_1, n''_2, \cdots, \sum_a n'_a)}
\end{equation}
have the same central charge. 
\begin{figure}
\center
\includegraphics[width=3.5in]{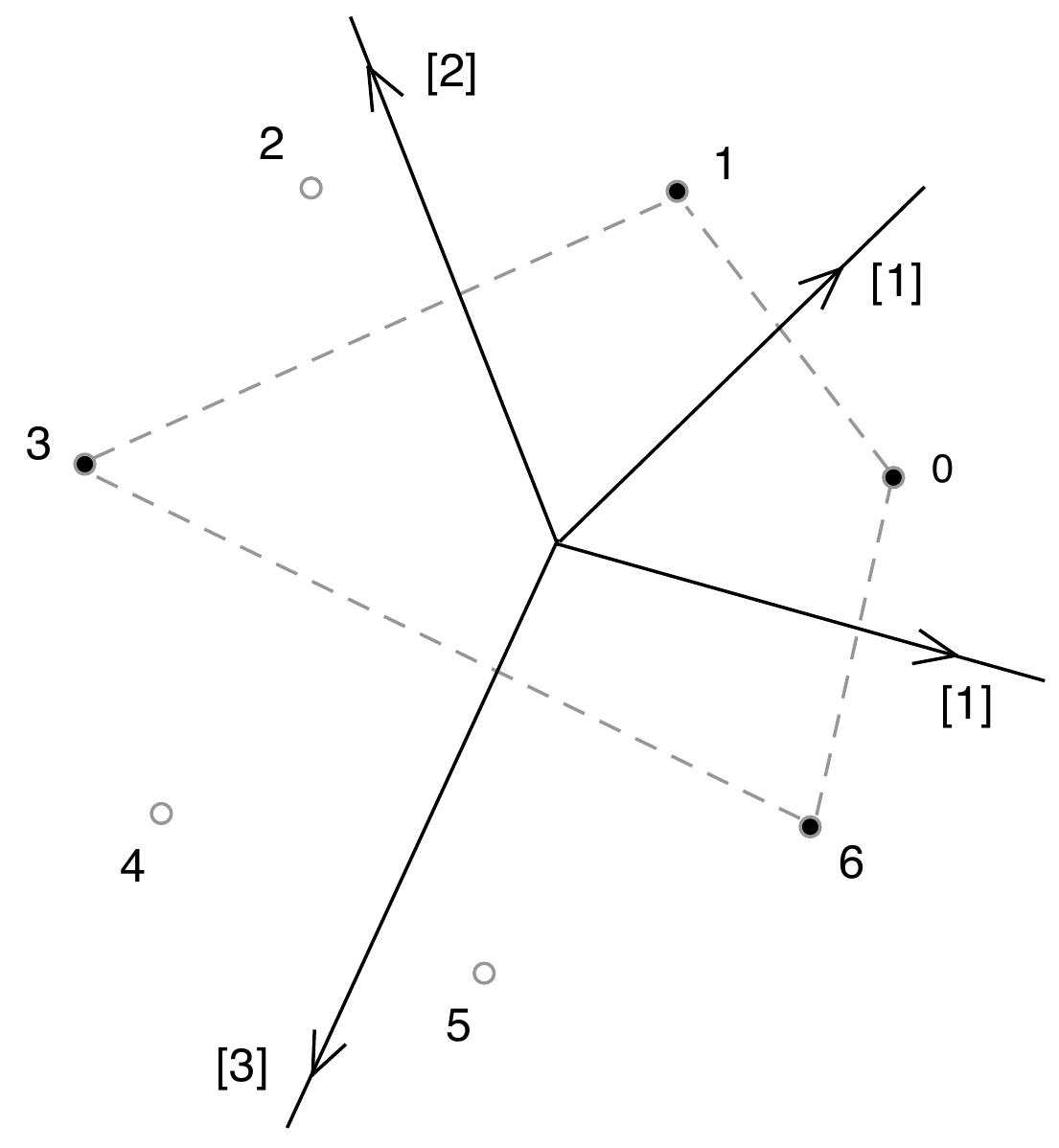}
\caption{A junction of domain walls between the first, third, sixth and seventh vacua of $\CN=1$ $SU(7)$ gauge theory.
This junction can be labelled by the cyclic sequence $2,3,1,1$ and is associated to the theory $T_{2,3,1,1}$. This junction can be resolved in two different ways, as depicted in figure \ref{fig:45} }
\label{fig:three}
\end{figure}
\begin{figure}
\center
\includegraphics[width=3in]{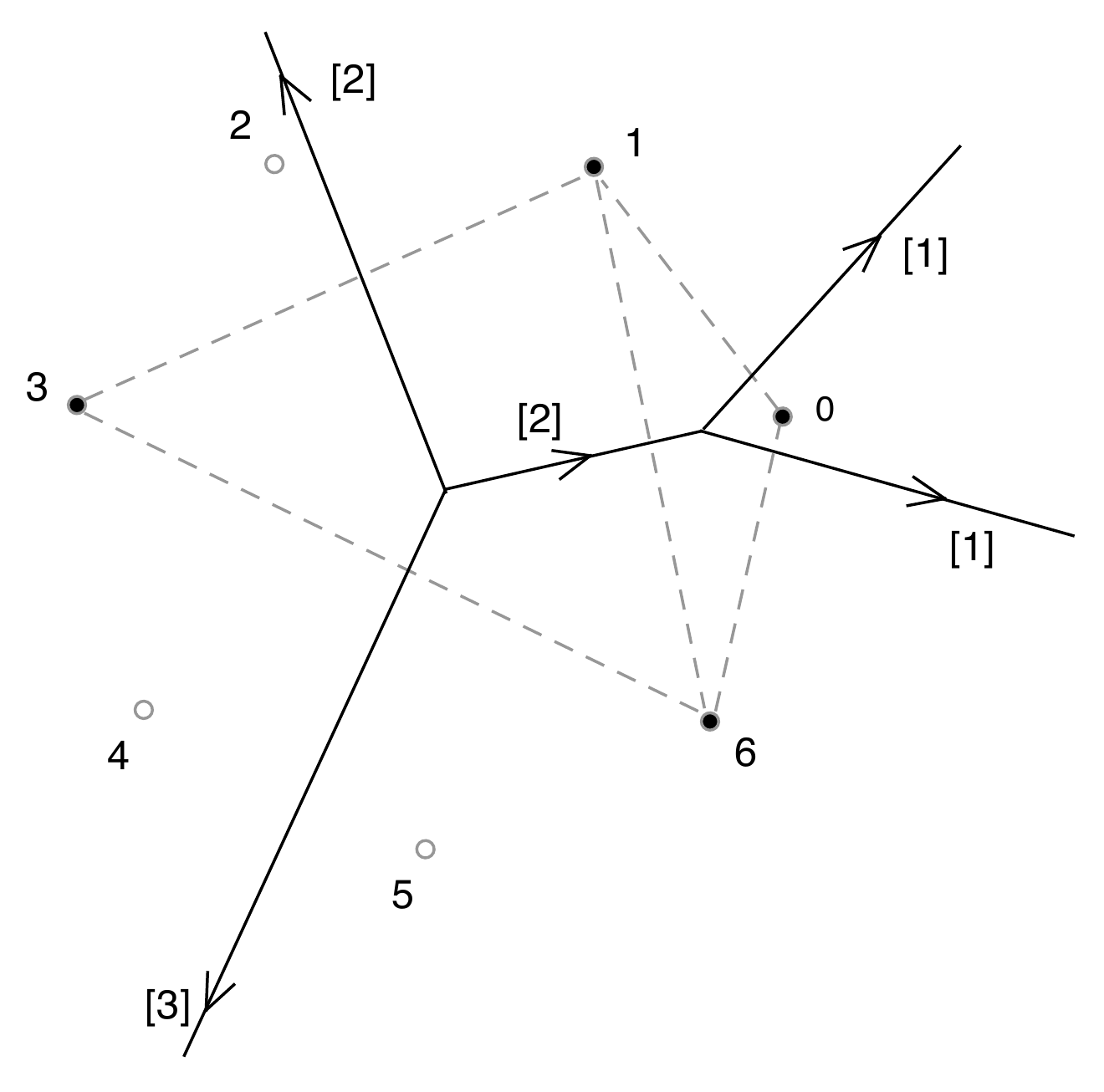}
\includegraphics[width=2.5in]{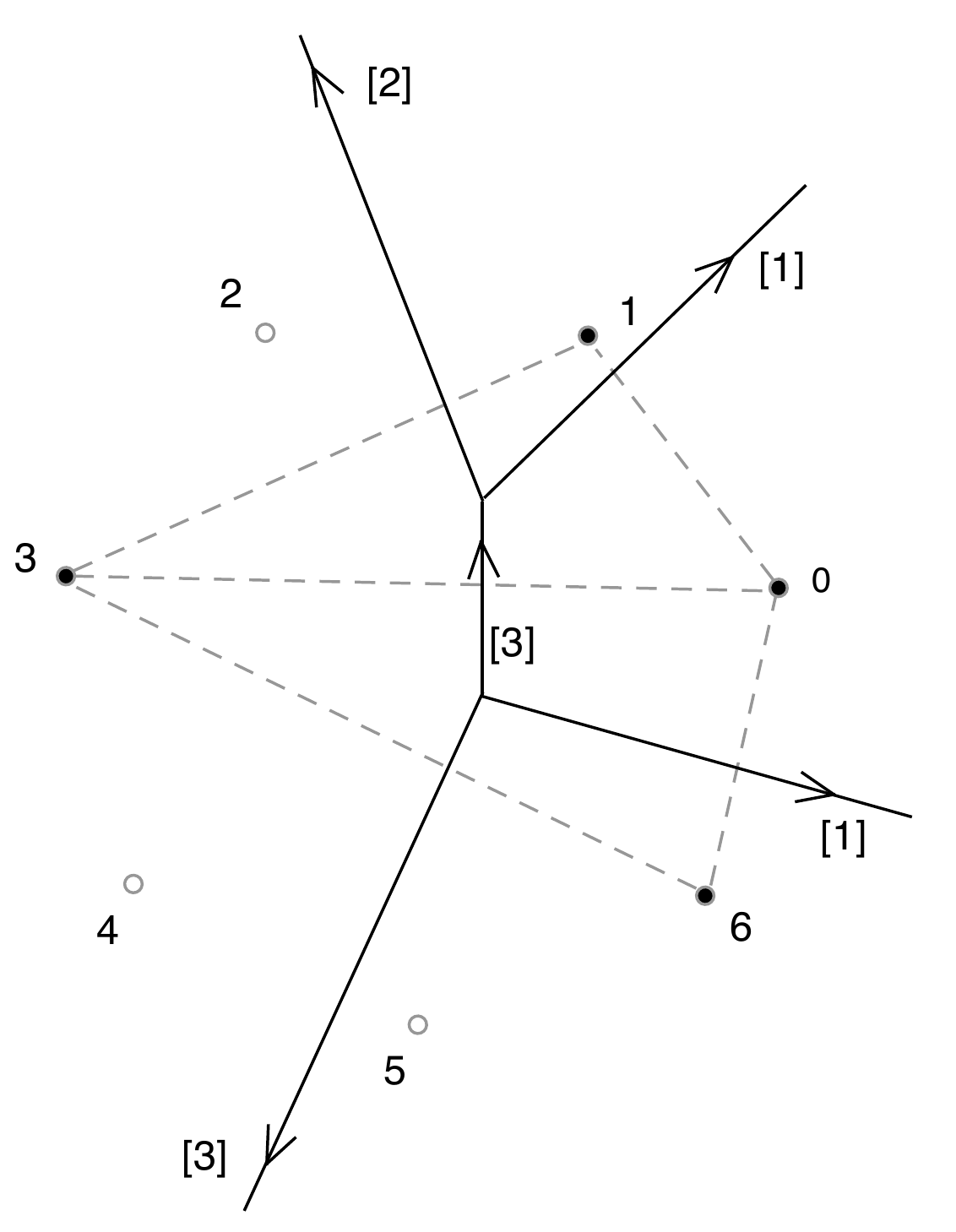}
\caption{The two possible resolutions of the $2,3,1,1$ junction. The first resolution is associated to the product theory $T_{1,1,5}\times T_{2,2,3}$, coupled through the 
$U(5)_7 \simeq U(2)_7$ super-Chern-Simons theory on the intermediate wall segment. The second is associated to the product theory $T_{1,2,4}\times T_{1,3,3}$, coupled through the 
$U(4)_7 \simeq U(3)_7$ super-Chern-Simons theory on the intermediate wall segment}
\label{fig:45}
\end{figure}

The relation holds thanks to a crucial observation: the product theory $[u(n)]_N \times [u(N-n)]_N$
has the same central charge as the $[u(N)_N]$ theory. Thus at the level of central charges, the factors 
of $[u(\sum_a n'_a)]_N$ and $[u(\sum_a n''_a)]_N$ in the denominators of the cosets cancel out against one numerator, 
and the product theory has the same central charge as the original junction theory. 

Digging a bit deeper, we find a much closer relation between $T_{(n_a)}$ and the product theory $T_{(n'_a, \sum_b n''_b)} \times T_{(n''_c, \sum_d n'_d)}$. 
First, we should observe a basic relation between $[u(n)]_N \times [u(N-n)]_N$ and the $[u(N)]_N$ theories, which 
is crucial in the self-consistency of the whole setup. Indeed, 
a domain wall between the $\ell$-th and the $(\ell+n)$-th vacua is the same, up to reversing orientation, as 
a domain wall between the $(\ell+n)$-th and the $(\ell+N)$-th vacua, as the vacua are cyclically ordered. 
The proposal of \cite{Acharya:2001dz} is consistent with this cyclic identification thanks to a level-rank duality $n \to N-n$
for the supersymmetric $U(n)_N$ Chern-Simons theory. This level-rank duality 
is encoded by the pairing of $[u(n)]_N$ and $[u(N-n)]_N$ primaries in the 
trivial supercoset
\begin{equation}
\left[\frac{u(N)_N}{u(n)_N \times u(N-n)_N} \right]
\end{equation}
which shows how the same supersymmetric $U(n)_N$ Chern-Simons theory could be coupled with the same orientation of the boundary
either to a chiral $[u(n)]_N$ model or to an anti-chiral $[u(N-n)]_N$ model. 
Equivalently,
\begin{equation}
[u(N-n)]_N = \left[\frac{u(N)_N}{u(n)_N} \right]
\end{equation}

Thus we can rewrite the product theory as
\begin{equation}
\left[\frac{u( \sum_a n'_a)_N}{\prod_a u(n'_a)_N}\right] \times \left[\frac{u(N)_N}{u( \sum_a n'_a)_N\times \prod_a u(n''_a)_N}\right]
\end{equation}
We can now see the full relation between the original junction theory and the theory on the split junctions: 
every primary in $T_{(n_a)}$ can be written as a sum of products of primaries in $T_{(n'_1, n'_2, \cdots, \sum_a n''_a)}$ and $T_{(n''_1, n''_2, \cdots, \sum_a n'_a)}$.
When the junction is split, the two factors in each primary move away from each other, but remain connected by the appropriate 
line defect of the $U(\sum_a n'_a)_N$ supersymmetric Chern-Simons theory living on the wall segment connecting the two new junctions.

\section{Super WZW theories and super-cosets} \label{sec:two}
We refer the reader to \cite{Kazama:1988uz} and references therein for material on supercosets such as $T_{n_a}$. 
We review here some basic facts necessary for our analysis. 

The supersymmetric affine current algebra $[{g}]_k$ at level $k$ for some simple Lie algebra ${g}$ is generated by 
adjoint free fermions $\psi^a$ together with a bosonic affine current algebra $[\hat g]_{\hat k}$ at level $\hat k = k-c_2({g})$. \footnote{ 
This is defined through the structure constants $f_{abc}$ by 
\begin{equation}
c_2({g}) \delta_{ab} = f_{acd} f_{bcd}
\end{equation}}
It is conventional to normalize the OPE of the free fermions and the bosonic currents as
\begin{align}
\psi_a(z) \psi_b(0) &\sim \frac{k}{2} \frac{\delta_{ab}}{z} \cr
\hat J_a(z) \hat J_b(0) &\sim \frac{\hat k}{2} \frac{\delta_{ab}}{z^2} + \frac{i f_{abc} \hat J_c(0)}{z}
\end{align}

The total level $k$ currents 
\begin{equation}
J_a = \hat J_a - \frac{i}{k} f_{abc} \psi_b \psi_c
\end{equation}
are super-partners of the free fermions under the super-current 
\begin{equation}
G = \frac{2}{k} \psi_a \hat J_a - \frac{2 i}{3 k^2} f_{abc} \psi_a \psi_b \psi_c
\end{equation}
The stress tensor is 
\begin{equation}
T = \frac{1}{k} :J_a J_a: - \frac{1}{k} \psi_a \partial \psi_a
\end{equation}
and the central charge 
\begin{equation}
c_{{g},k} = \left( \frac{\hat k}{k} + \frac{1}{2} \right)  \mathrm{dim}\, {g}
\end{equation}

The primary operators in the NS sector coincide with the primary operators of the bosonic WZW model.
All other operators are descendants under the action of $\psi_a$ and $\hat J_a$. Thus the primaries are labelled by a 
representation $r$ of ${g}$ in the appropriate range determined by the bosonic level $\hat k$. 
They have conformal dimension 
\begin{equation}
\Delta_r = \frac{c_2(r)}{k}
\end{equation}

The $[u(n)]_N$ theory at level $N$ can be built combining an $[su(n)]_N$ theory at level $N$ 
by and an $[u(1)]$ theory at appropriate level. Thus we add to the $[su(n)]$ fields a
free fermion $\psi_{U(1)}$ and affine current $\hat J_{U(1)}$
and dress up the $[su(n)]$ primaries with appropriate vertex operators for the $\hat J_{U(1)}$
current algebra, so that they lie in $u(n)$ representations. 
The extra $U(1)$ vertex operator shifts the conformal dimension of an operator labelled by a Young tableau
with $m$ boxes by $\frac{m^2}{2 n k}$. For example, the dimensions of 
the primary in the fundamental, $m$-th symmetric and $m$-th antisymmetric representations become respectively
\begin{equation}
\frac{n}{2 k} \qquad \qquad \frac{m(n+m-1)}{2k} \qquad \qquad \frac{m(n-m+1)}{2k}
\end{equation}
The central charge of the $[u(n)]_N$ theory is
\begin{equation}
c_{n,N} =  \left(  \frac{3}{2} - \frac{n}{N} \right) (n^2-1) + \frac{3}{2}
\end{equation}

\subsection{Super-cosets}
If we have a sub-algebra ${h} \in {g}$, the fermions 
$\psi_a^H$ and total currents $J_a^H$ which lie in the sub-algebra define an 
super affine $[{h}]$ current algebra. The bosonic part of this super affine algebra 
is ${\it not}$ given by the corresponding subset of the $\hat J_a$ currents: 
it receives an extra contribution from the free fermions which lie in the complement ${g}/{h}$. 
Correspondingly, the super-coset $[\frac{g}{h}]$ differs from the corresponding bosonic coset theory by the contributions of these fermions.

We are particularly interested in super-cosets where the bosonic level of the $[{g}]$ current algebra is zero. 
This means that the coset primary fields are obtained by decomposing polynomials of the adjoint free fermions 
$\psi_a$ into products of $[{h}]$ and $[\frac{g}{h}]$ primary fields and descendants. As the 
fermions which lie in the ${h}$ sub-algebra are part of the current algebra of the $[{h}]$ theory, we can restrict ourselves to 
polynomials in the fermions which lie in the complement ${g}/{h}$. 

We can now specialize to the case of interested for this paper, which are cosets of the form 
\begin{equation}
T_{(n_a)}= \left[ \frac{u(N)_N}{\prod_a u(n_a)_N} \right]
\end{equation}
We can start with the case with two denominator factors, $n_1=n$ and $n_2 = N-n$. 
The central charges of $[u(n)]_N$ and $[u(N-n)]_N$ add up to the central charge of $[u(N)]_N$.
\begin{equation}
c_{n,N} + c_{N-n,N} = c_{N,N}
\end{equation}

Furthermore, the complement $u(N)/\left(u(n) \times u(N-n) \right)$ consists of the product of 
a fundamental representation of $u(n)$ and anti-fundamental of $u(N-n)$.
The sum of conformal dimensions of a $[u(n)]_N$ primary in the fundamental representation 
and an $[u(N-n)]_N$ primary in the anti-fundamental representation is exactly $1/2$. 
Thus we can directly decompose the fermions in the complement $u(N)/\left(u(n) \times u(N-n) \right)$ as a product of (anti)fundamental primaries of $[u(n)]_N$ and $[u(N-n)]_N$:
\begin{equation}
\psi_{i,i'} = \phi_i^{u(n)} \otimes \phi_{i'}^{u(N-n)}
\end{equation}

In a similar fashion, we can decompose the product of $2$ fermions
\begin{equation}
\psi_{i,i'} \psi_{j,j'}= \phi_{[ij]}^{u(n)} \otimes \phi_{(i'j')}^{u(N-n)} + \phi_{(ij)}^{u(n)} \otimes \phi_{[i'j']}^{u(N-n)} 
\end{equation}
Indeed, the sum of conformal dimensions of a two-index symmetric tensor in $[u(n)]_N$ and a two-index anti-symmetric tensor 
in $[u(N-n)]_N$ is exactly $1$, and the same is true for the opposite combination. 
In general, all the pairs of representations which we find in the product of $m$ fermions have dimensions which add up to $m/2$.

This verifies that the $T_{n,N-n}$ coset is trivial, and that we can identify
\begin{equation}
[u(N-n)]_N = \left[\frac{u(N)_N}{u(n)_N} \right]
\end{equation}
This is the level-rank duality of supersymmetric Chern Simons theory.
Notice that the supersymmetric level-rank duality reduces immediately to the usual non-supersymmetric statement: the 
bosonic $[\hat u(n)]_{N-n}$ and $[\hat u(N-n)]_n$ WZW theories can be combined into a set of $n(N-n)$ free fermions.
 
This analysis makes rather clear the structure of the more general coset $T_{(n_a)}$: we have a set of bifundamental 
fermions for each pair of $u(n_a)$ and $u(n_b)$ subalgebras, which can be decomposed into a pair of bosonic 
WZW theories $[\hat u(n_a)]_{n_b}$ and $[\hat u(n_b)]_{n_a}$. Thus the primary fields of the supercoset can be 
decomposed into products of primaries of the bosonic cosets 
\begin{equation}
\prod_a \left[\frac{\prod_{b \neq a} \hat u(n_a)_{n_b}}{\hat u(n_a)_{N-n_a}} \right]
\end{equation}
with appropriate pairing between the representation labels of $[\hat u(n_a)]_{n_b}$ and $[\hat u(n_b)]_{n_a}$. 

\subsection{Line operators and junctions}
The most important observables in Chern-Simons theories are Wilson line operators. The structure of our 
conjectural junction SCFT allows the Wilson line operators of the domain wall Chern-Simons theories to end 
on operators at the junction, but only according to certain somewhat complicated patterns. For example, a 
Wilson loop in the fundamental representation of an $U(n_a)$ theory will have to end on a junction operator which arises from 
one of the groups of $n_a \times n_b$ bifundamental fermions for some $n_b$. That operator must also be attached to an 
anti-fundamental Wilson loop for the corresponding $U(n_b)$ CS theory.

This is a complicated way to say that a fundamental Wilson line in one of the domain walls cannot end, but must continue 
as a fundamental Wilson lines in another of the domain walls ending at the junction. 
Wilson lines in higher representations may split and recombine in various ways as they cross the junction. 

We propose a simple explanation of this phenomenon: the Wilson line operators of the domain wall Chern-Simons theories
arise in the UV from elementary Wilson line operators of the $\CN=1$ four-dimensional Yang-Mills theory. 
We are used to Wilson line operators being confined in the bulk of the theory, because at low energy they 
must have confining strings ending on them. On the other hand, it is known that confining strings can end on domain walls
as well. Thus if we bring a Wilson line operator from the bulk to a domain wall, the confining string may detach 
and move away along the domain wall, leaving the Wilson line operator behind. 

Another useful point of view \cite{Witten:1997ep,Rey} is that different vacua of the theory are associated to the condensation of monopoles/dyons with distinct electric charge. 
Thus the electric flux sourced by a Wilson line operator living at a domain wall could be screened by simultaneous condensation of 
dyons on the two sides of the wall, conspiring to cancel out the overall magnetic flux of the condensate and leaving an electric flux to screen the Wilson line operator. In this picture, the Wilson line operator would be accompanied by a fixed amount of magnetic flux 
threading the domain wall. This magnetic flux will change the mutual statistics of domain wall Wilson lines and may explain why do they behave as Wilson lines in a Chern-Simons theory. See also \cite{2013arXiv1302.6535M} for a recent discussion of similar facts in a condensed matter setting, which provided the initial motivation for this paper.  

We will now give a concrete realization of this idea in a convenient setup: weakly mass deformed $\CN=2$ gauge theory, where 
monopole and dyon condensation can be made very concrete. 

\section{Domain walls in mass-deformed $\CN=2$ gauge theory} \label{sec:three}
We can start with the simplest possible example: pure $SU(2)$ gauge theory. 
The $\CN=1$ mass-deformation of this theory is well understood \cite{Seiberg:1994rs}. The two vacua of the theory are localized at the monopole and 
dyon points of the Coulomb branch respectively,
where the added $\CN=1$ superpotential induces a condensation of the BPS monopole or dyon
which become massless at these loci. 
The monopole condensation breaks the low-energy $U(1)$ gauge theory and produces a mass gap. 

A domain wall between the two vacua can be thought of as a path in the Coulomb branch joining the two vacua. 
Far from the monopole or dyon points, where we can ignore the effects of the BPS particles, the path 
should follow a trajectory of constant phase for the superpotential, i.e. for the standard Coulomb branch parameter $u$. 
Near the monopole or dyon points one can include the appropriate BPS particles and solve numerically the BPS domain wall equations 
for the Coulomb branch parameter and the monopole or dyon webs. The three solutions can be matched together into a single flow \cite{Kaplunovsky:1998vt}.

Here we are mainly concerned with a simple observation. The three phases of the flow along the Coulomb branch have the following features.
Near the monopole point, the monopole has a vev, which interpolates from the vacuum value at infinity to an exponentially small value. 
In a duality frame where the monopole is electrically charged, the gauge fields are Higgsed and very massive in the region where the
monopole vev is very large, but become very light in the transition region. At low energy, we can approximate this by 
removing the gauge fields from the region far from the domain wall and setting Dirichlet boundary conditions for them at the boundary of the transition region. 
Near the dyon point, a similar reasoning applies, but in a different duality frame, where the dyon is electrically charged. 

We can thus approximate the system by a $U(1)$ gauge theory on a segment, with boundary conditions at the two ends which are related to Dirichlet 
by distinct electric-magnetic duality transformations. There are many equivalent ways to treat this system. The most intuitive is to work in the standard electric-magnetic duality frame 
for the semi-classical theory, where the W-boson has electric charge $2$, the monopole has magnetic charge $1$ and the dyon has magnetic charge $1$, electric charge $-2$. Then the boundary condition at
the monopole point becomes a Neumann boundary condition. The boundary condition at the dyon point becomes a deformed Neumann boundary condition, with two units of Chern-Simons coupling 
added at the boundary. \footnote{This $\CN=2$ field theory model seems a convenient setup where to fix the relative orientation of the Chern-Simons coupling and the BPS junctions}. Thus the gauge theory on the segment reduces to a supersymmetric $U(1)_2$ Chern-Simons theory with level $2$, as expected!

We can see very concretely how bulk line operators are related to the Chern-Simons theory Wilson line operators. 
It is easier to start with line operators for the low energy Abelian $\CN=2$ gauge theory. 
The full non-Abelian line defects in the UV theory decompose into sums of line defects of the Abelian theory 
in a predictable way \cite{Gaiotto:2010be}, and the analysis below can be applied to each IR summand. 

The monopole and dyon condensates on either sides of the domain wall
can combine to screen any charge of the form $(a + b, -2b)$. The condensate can screen charges with any $a$, $b$, not necessarily integral. 
Thus every Abelian line defect could live at the domain wall, without requiring a bulk confining string. 
On the other hand, charges with integral $a$ and $b$ can be screened by actual monopole and dyon particles, and 
thus should become trivial in the IR. The set of interesting domain wall line defects is given by 
the charge lattice modulo the lattice of screening charges and consists of a single non-trivial element, a 
Wilson loop of electric charge $1$. This clearly maps to the basic Wilson loop of the $U(1)_2$ Chern-Simons theory. 
The screening condensate threads half a unit of magnetic flux through the domain wall Wilson loop. 
This gives an alternative way to understand how the domain wall Wilson loop acquires the 
anyonic properties of the $U(1)_2$ Chern-Simons Wilson loop. 

It is clear that the discussion above applies equally well to domain walls in any weakly mass-deformed $\CN=2$ gauge theory. 
Simple domain walls will correspond to certain paths in the Coulomb branch, joining singularities where a maximal number of mutually local particles become massless. 
Given an electric-magnetic duality transformation $g$ between the corresponding duality frames where these particles are electrically charged, 
we can approximate the system by putting the bulk Abelian gauge theory on a segment, with Dirichlet b.c. on the two ends and a duality wall \cite{Gaiotto:2008ak}
for the $g$ transformation in the middle. This will produce an Abelian Chern-Simons gauge theory which depends on $g$ up to multiplication on the left or the right 
by transformations which preserve Dirichlet b.c., which are the same that leave the condensing particles electrically charged. 

This line of reasoning shows that basic domain walls in a weakly mass-deformed $\CN=2$ gauge theory should support a very specific Abelian Chern-Simons TFTs 
in the infrared, determined by the electric-duality transformations encountered by the flow along the Coulomb branch.
 Conversely, a domain wall can support more complicated IR TFTs only if the description as a flow between singularities in the Coulomb branch 
is invalid. An example might be a domain wall which flows out of a singularity, passes ``through'' a second singular locus (i.e. is described by a flow in the 
full local effective field theory at that point) and then ends at a third singular point.

This analysis is in obvious tension with our assumption that a non-Abelian Chern-Simons theory is supported on the domain walls of pure $SU(N)$ 
$\CN=1$ gauge theory. There are at least two possible ways to resolve this discrepancy without renouncing our assumption. 
The first possibility is that the domain walls with $n>1$ simply do not appear in the weakly mass-deformed $\CN=2$ gauge theory.
They may literally not exist, or they may be realized by flows which pass through extra singular points. 
Conversely, if we assume that the gauge charges of the massless particles at the two endpoints of the flow generate the full lattice of charges populated by BPS particles of the theory,
then the lattice of charges of line defects modulo the lattice of charges generated by the massless particles
is $Z_N$. This agrees with the $U(1)_N$ Chern-Simons theory we expect on $n=1$ domain walls. 

The second possibility is that the $n>1$ domain walls exist and are simple in the weakly mass-deformed $\CN=2$ gauge theory,
and a phase transition occurs as we tune the mass deformation to be large in order to flow to the $\CN=1$ gauge theory. 
The domain wall trajectories on the Coulomb branch depend on the metric on the Coulomb branch fields, 
which will in general be affected by a non-small mass deformation. If the trajectory is deformed across singularities in the Coulomb branch, 
the e.m. transformation $g$ and the candidate Abelian CS theory will change. This can be thought of as a phase transition, due to a 
bulk BPS particle becoming massless on the wall. When the mass deformation become sufficiently large, the picture of a specific trajectory on the Coulomb branch 
loses validity, and we can imagine the domain wall theory may undergo further phase transitions to the desired non-Abelian Chern-Simons theories. 

The second possibility offers an important cautionary tale. It is perfectly possible that similar phase transitions would occur as we 
deform the $\CN=1$ gauge theory to the brane systems which are used to derive our main assumption on the domain walls worldvolume theories. 
The brane description is not obviously more trustworthy than the weakly mass-deformed $\CN=2$ description.
It would be interesting to discriminate between the two possibilities through a concrete numerical investigation of solutions of the BPS domain wall equations in the 
Coulomb branch of the pure $SU(N)$ Seiberg-Witten theory. We leave this analysis to future work.

\section{Generalizations and open questions}
In the main text we encountered two calculations which may further test our proposal: 
fixing the absolute sign of the domain wall Chern-Simons couplings, either through string theory 
or through the weakly mass-deformed $\CN=2$ theory and exploring numerically the BPS domain walls in the weakly mass-deformed 
$\CN=2$ theory. We can add to these a third test: compute from first principles the gravitational anomaly of a domain wall junction
and compare it with the central charge of $T_{(n_a)}$.

A broader open question is what is the correct generalization of this work to 
other bulk gauge groups $G$, which are expected to have $c_2(G)$ vacua. 
To the best of our knowledge there is no concrete proposal for the 
3d TFTs living on the domain walls of these theories, only a proposal for their Witten index \cite{Acharya:2001dz}.

There are two other obvious classes of coset models which can satisfy junction splitting conditions, 
due to the level-rank dualities of orthogonal and symplectic WZW models:
\begin{equation}
\left[ \frac{sp(N)_{N+1}}{\prod_a sp(n_a)_{N+1}} \right] \qquad \qquad \left[ \frac{so(N)_{N-2}}{\prod_a so(n_a)_{N-2}} \right]
\end{equation}
It is not straightforward, though, to match these two series to some choice of bulk gauge group, though the former
is a reasonably good fit for an $Sp(N+1)$ bulk gauge group. 

Another interesting extension of this work is to study the interplay with the precise choice of bulk gauge group, say for example $PSU(N)$ instead of $SU(N)$. Such choices would presumably affect the domain walls 3d TFTs, 
as they modify the choices of available line defects for the theory. They may also change the topological nature 
of the bulk vacua \cite{Aharony:2013hda}. Although we would not expect these choices to change the dynamics of the junction 
theories, they may affect in interesting ways the choice of available operators in the theory. 

Finally, one may ask similar questions about other gapped $\CN=1$ gauge theories. 

\section*{Acknowledgements}
We are grateful to the organizers of the {\it Emergence and Entanglement II} conference, which provided 
the initial inspiration for this work. 
The research of DG was supported by the Perimeter Institute for Theoretical Physics. Research at Perimeter Institute is supported by the Government of Canada through Industry Canada and by the Province of Ontario through the Ministry of Economic Development and Innovation.

\bibliographystyle{JHEP_TD}
\bibliography{N1webs}

\end{document}